\documentclass[aps,pra,reprint,superscriptaddress,showpacs]{revtex4-1}

\usepackage{graphicx}
\usepackage{amssymb,amsmath}
\usepackage{bm}

\begin{document}

\title{Quantum-defect theory of resonant charge exchange}

\author{Ming Li}
\email[]{ming.li3@rockets.utoledo.edu}
\author{Bo Gao}
\email[]{bo.gao@utoledo.edu}
\affiliation{Department of Physics and Astronomy,
    University of Toledo, Mailstop 111,
    Toledo, Ohio 43606,
    USA}
\affiliation{State Key Laboratory of Low-Dimensional Quantum Physics, 
	Department of Physics, Tsinghua University, 
	Beijing 100084, 
	China}

\date{May 22, 2012}

\begin{abstract}

We apply the quantum-defect theory for $-1/R^4$ potential to study
the resonant charge exchange process. We show that by taking
advantage of the partial-wave-insensitive nature of the formulation,
resonant charge exchange of the type of $^1$S+$^2$S can be accurately
described over a wide range of energies using only three parameters,
such as the \textit{gerade} and the \textit{ungerade} $s$ wave scattering lengths, and the
atomic polarizability, even at energies where many partial waves
contribute to the cross sections. The parameters can be determined
experimentally, without having to rely on accurate potential energy
surfaces, of which few exist for ion-atom systems. The theory
further relates ultracold interactions to interactions at much higher
temperatures.

\end{abstract}


\maketitle

\section{Introduction}

With the recent emergence of cold-ion experiments, either with 
trapped ions \cite{gri09,zip10,zip10b,rel11,hal11} or in the context of
cold plasmas \cite{kil07},
there is a growing interest in 
ion-atom interactions at cold temperatures \cite{idz09,gao10a,idz11},
including, in particular, the resonant charge exchange process
(see, e.g., Refs.~\cite{cot00,Bodo2008,zha09}), such as
\[
\mathrm{Na}^++\mathrm{Na}\longrightarrow \mathrm{Na}+\mathrm{Na}^+ \;.
\]
Despite being one of the simplest reactive processes that has been a subject
of study for a long time \cite{mot65}, 
{\em quantitative} understanding of resonant  
charge exchange remains difficult, especially at cold temperatures.
This difficulty stems from its sensitive dependence on the potential
energy surfaces (PES), a common difficulty shared by all heavy
particle interactions at cold temperatures (see, e.g., \cite{gao96}),
including not only ion-atom interaction, 
but also atom-atom interactions \cite{chi10},
and chemical reactions whenever the Langevin assumption breaks
down \cite{gao10b,gao11a}.

In the case of atom-atom interaction, this difficulty has only been overcome
by incorporating a substantial amount of spectroscopic data, 
especially data close to the dissociation limit, 
to fine tune the PES (see, e.g., Ref.~\cite{Ferber2009,kno11}).
Without such fine tuning, no \textit{ab initio}
PES for alkali-metal systems has been sufficiently accurate to predict 
the scattering length and other scattering characteristics around the
threshold.
The availability of such data, however, is limited mostly to
alkali-metal atoms and a few other species that can be cooled.
For ion-atom systems, with a few exceptions that came close \cite{hec02},
no such data are yet available, though recent efforts on the 
trapping and cooling of molecular ions 
(see, e.g., Refs.~\cite{HudsonPRA2009,NguyenNJP2011,NguyenPRA2011,wil12})
show considerable future promise.
This status on the ion-atom PES is such that
at the moment, with the possible exception of H$^+$+H and 
its isotopic
variations \cite{Bodo2008,iga99,esr00}, no other predictions for cold
or ultracold ion-atom processes, including resonant
charge exchange, can yet be trusted before experimental
verification.

We present here a quantum-defect theory (QDT), 
not only as a general approach to ion-atom interactions, 
but also as one specific method of dealing 
with this difficulty of sensitive dependence on PES.
It is an initial application of the 
QDT for $-1/R^4$ potential \cite{wat80,fab86,gao10a},
as formulated in Ref.~\cite{gao10a}, to the resonant charge
exchange process. We show that by taking advantage of the 
partial-wave-insensitive nature of the QDT formulation \cite{gao01,gao08a,gao10a},
resonant charge exchange of the type of $^1$S+$^2$S,
applicable to Group IA (alkali), Group IIA (alkaline earth), and helium 
atoms in their ground states, can be accurately
described over a wide range of energies using only three parameters,
such as the \textit{gerade} and the \textit{ungerade} $s$ wave scattering lengths, and the
atomic polarizability, even at energies where many partial waves
contribute to the cross sections. The parameters can be determined
experimentally, without having to rely on accurate PESs. 
The theory further relates ultracold ion-atom interactions to 
interactions at much higher temperatures.
It is the beginning of a much broader program aimed at
connecting ultracold interactions and
reactions of all types, to their behaviors at much 
higher energies and/or temperatures that are more relevant
to astrophysics and everyday chemistry.

For simplicity and for purposes of making connections with 
existing formulations and setting benchmarks for further 
theoretical developments, we adopt here the so-called 
elastic approximation (see, e.g., Refs.~\cite{cot00,zha09}) which ignores 
the hyperfine and isotope effects. 
Going beyond this approximation will require
a multichannel quantum-defect theory (MQDT)
formulation of near-resonant charge exchange,
along the lines of Refs.~\cite{gao96,gao05a} and similar in spirit
to recent works of Idziaszek \textit{et al.} \cite{idz09,idz11}.
Such a theory will be better understood after
the conceptual and numerical developments of this
work, which relies only on the single-channel QDT \cite{gao10a}.
The multichannel formulation will be presented in 
a future publication.

The paper is organized as follows. In Sec.~\ref{sec:qdt},
we outline the application of QDT to the resonant charge
exchange of the type of $^1$S+$^2$S.
Section~\ref{sec:3pqdt} discusses two equivalent
three-parameter descriptions derived from the QDT formulation.
In Sec.~\ref{sec:Na}, the three-parameter QDT description is
thoroughly tested for the case of resonant charge exchange
of $^{23}$Na, both through comparison with previous
results of C\^ot\'e and Dalgarno \cite{cot00},
and by comparing our own numerical results with the corresponding
QDT parametrization results.
We conclude in Sec.~\ref{sec:concl} with further discussions on
the implications of our results in the more general context of
ion-atom interactions.

\section{QDT of resonant charge exchange}

\subsection{General considerations for $^1$S+$^2$S type of systems}
\label{sec:qdt}

Consider the system of an atom and its ion, one in a $^1$S state, one
in a $^2$S state. This type covers resonant charge exchange of both
Group IA (alkali), Group IIA (alkaline earth), and helium atoms in their
ground states. 
Such a system correlates to two Born-Oppenheimer molecular states, characterized
by $^2\Sigma_{g,u}^+$. The other, energetically higher, electronic states can be
ignored for collision energies much smaller than 
the first electronic excitation energy \cite{mot65}. 
In the case of exact resonance or in the elastic approximation 
which ignores hyperfine structures 
and/or the isotope shifts, the two coupled radial Schr\"{o}dinger equations
in the atomic basis become decoupled in the molecular basis,
reducing the understanding of resonant charge exchange to two
single-channel equations for the \textit{gerade}, $g$, and the \textit{ungerade},
$u$, states, respectively \cite{mot65,cot00,zha09}
\begin{equation}
\left[
-\frac{\hbar^2}{2\mu}\frac{d^2}{dR^2}+\frac{\hbar^2l(l+1)}{2\mu
R^2}+V_{g,u}(R)-\epsilon \right]u_{\epsilon l}^{g,u}(R)=0 \;.
\label{eq:rschr}
\end{equation}
Here $\mu$ is the reduced mass, $\epsilon$ is the energy in
the center-of-mass frame, $V_{g,u}(R)$ represent the two
potential energy curves for the \textit{gerade} and the
\textit{ungerade} states, and
$u_{\epsilon l}^{g,u}(R)$ are the corresponding 
radial wave functions for the $l$th
partial wave.

In terms of the two phase shifts, $\delta^{g,u}_l$, for the 
\textit{gerade} and \textit{ungerade} states in partial wave $l$, 
as determined from the solutions
of Eq.~(\ref{eq:rschr}), the total charge exchange 
cross section $\sigma_{\mathrm{ex}}$ can be written as \cite{mot65,cot00,zha09}
\begin{equation}
\sigma_{\mathrm{ex}}(\epsilon) =\frac{\pi}{k^2}\sum_{l=0}^{\infty}
    (2l+1)\sin^2(\delta^{g}_{l}-\delta^{u}_{l}) \;.
\label{eq:sigmaex}
\end{equation}
It contains the physical concept that resonant charge exchange
is due to the phase difference between the \textit{gerade} and the \textit{ungerade}
molecular states. For elastic and total cross sections,
it is convenient to first define two single-channel 
``molecular'' cross sections
\begin{equation}
\sigma^{g,u} =
\frac{4\pi}{k^2}\sum^\infty_{l=0}(2l+1)\sin^2(\delta^{g,u}_l) \;,
\label{eq:sigmol}
\end{equation}
in terms of which the total cross section is given by
$\sigma_\text{tot} = (\sigma^g+\sigma^u)/2$,
and the elastic cross section is given by
$\sigma_\text{el} = \sigma_\text{tot}-\sigma_\text{ex}$ \cite{mot65,cot00,zha09}.

For a $^1$S+$^2$S type of system with either a $^1$S or a $^2$S atom 
in its ground state, the potentials
$V_{g,u}(r)$ in Eq.~(\ref{eq:rschr}) have the same leading term
$-C_4/R^4$ at long range, where $C_4>0$ is given in atomic units
by $C_4=\alpha_1/2$ with $\alpha_1$ being the static dipole 
polarizability of the atom.
Application of the single channel QDT for $-1/R^4$
type of potential \cite{gao10a} gives an efficient characterization
of the phase shifts $\delta^{g,u}_l$, leading to an efficient
characterization of the resonant charge exchange process.

Specifically, for a potential with a long-range behavior
of $V\sim -C_n/R^n$, the tangent of the phase shift is given
in QDT by \cite{gao08a}
\begin{equation}
\tan\delta_l =
    \left(Z^{c(n)}_{gc}K^c -Z^{c(n)}_{fc}\right)
    \left(Z^{c(n)}_{fs}-Z^{c(n)}_{gs}K^c \right)^{-1} \;.
\label{eq:qdtpe}
\end{equation}
Here $K^c(\epsilon,l)$ is a dimensionless short-range 
$K^c$ matrix \cite{gao08a}.
$Z^{c(n)}_{xy}(\epsilon_s,l)$ are universal QDT functions 
for $-1/R^n$ type of potential. 
They are functions of the angular momentum $l$ and a
scaled energy $\epsilon_s = \epsilon/s_E$, where
$s_E = (\hbar^2/2\mu)(1/\beta_n)^2$ is the energy scale and 
$\beta_n = (2\mu C_n/\hbar^2)^{1/(n-2)}$
is the length scale for the $-C_n/R^n$ potential.
Explicit expressions for 
$Z^{c(n=4)}_{xy}$, applicable to the polarization potential,
are given in the Appendix. 
As explained in Ref.~\cite{gao08a}, Eq.~(\ref{eq:qdtpe})
includes not only the effect of long-range phase shift,
but also effects of quantum reflection and tunneling,
which are the key differences between long-range potentials
with $n>2$ and those with $n<2$.

All cross sections for resonant charge exchange 
can be written explicitly 
in terms of $\tan\delta^g_l$ and
$\tan\delta^u_l$ for the \textit{gerade} and the \textit{ungerade} states.
\begin{equation}
\sigma_{\mathrm{ex}}(\epsilon) =\frac{\pi}{k^2}\sum_{l=0}^{\infty}
    (2l+1)\frac{(\tan\delta^{g}_l-\tan\delta^{u}_l)^2}
    {(1+\tan^2\delta^{g}_l)(1+\tan^2\delta^{u}_l)}\;,
\end{equation}
and
\begin{equation}
\sigma^{g,u} =
\frac{4\pi}{k^2}\sum^\infty_{l=0}(2l+1)\frac{\tan^2(\delta^{g,u}_l)}
	{1+\tan^2(\delta^{g,u}_l)}\;.
\end{equation}
For sufficiently large $l$ and away from a shape resonance, 
$\tan\delta_l$ is independent of
the short-range parameter and is given, for both $g$ and $u$ states, by
the Born approximation (see, e.g., Ref.~\cite{lan77})
\begin{equation}
\tan\delta_l \sim \frac{\pi}{(2l+3)(2l+1)(2l-1)}\epsilon_s \;.
\end{equation}
This $1/l^3$ type of behavior for large $l$ ensures convergence 
in summations over $l$ in all total
cross section calculations.


The application of QDT allows the description of resonant charge
exchange in terms of the $C_4$ coefficient, equivalently the atomic
polarizability $\alpha_1$, and two short-range $K$ matrices,
$K^c_g(\epsilon,l)$ for the \textit{gerade} state and $K^c_u(\epsilon,l)$
for the \textit{ungerade} state. Such a description is exact if
the energy and the partial wave dependences of the $K^c$s are
fully accounted for.
More importantly, QDT allows for efficient parametrizations
of resonant charge exchange by taking advantage of the
fact that the short-range $K^c$ matrices depend not only weakly on energy,
but also weakly on the partial wave $l$ for both atom-atom
and ion-atom interactions \cite{gao01,gao08a,gao10a}.
Through an example of Na$^+$+Na, we show that even the
simplest parametrization, corresponding to ignoring
the $\epsilon$ and $l$ dependences of the $K^c$s completely,
provides an accurate description of resonant charge
exchange over a wide range of energies, including energies
at which tens of partial waves contribute.
 
\subsection{Three-parameter QDT descriptions}
\label{sec:3pqdt}

A three-parameter parametrization of resonant charge
exchange for a $^1$S+$^2$S system results from ignoring both
the energy and the partial wave dependences of 
$K^c_g(\epsilon,l)$ and $K^c_u(\epsilon,l)$.
Specifically, it corresponds to the approximation of
$K^c_g(\epsilon,l)\approx K^c_g(\epsilon=0,l=0)$
and $K^c_u(\epsilon,l)\approx K^c_u(\epsilon=0,l=0)$.
Using $K^c_g$ and $K^c_u$ as the shorthand notation for
the resulting constant $K^c$s, we have one of the three-parameter
parametrizations for resonant charge exchange, with
two short-range parameters $K^c_g$ and $K^c_u$, characterizing
the short-range ion-atom interaction,
and one long-range parameter, $C_4$ or the atomic 
polarizability $\alpha_1$, characterizing the strength of
the long-range interaction.

A mathematically equivalent three-parameter parametrization
is in terms of two $s$ wave scattering lengths, $a_{gl=0}$ 
and $a_{ul=0}$, for the \textit{gerade} and the \textit{ungerade} state, respectively,
and the atomic polarizability $\alpha_1$.
This is derived from the first parametrization by noting
that the $K^c(\epsilon=0,l=0)$, for both $g$ and $u$ states,
are related rigorously to the corresponding $s$ wave
scattering lengths by \cite{gao03,gao04b}
\begin{equation}
a_{l=0}/\beta_n = \left(b^{2b}\frac{\Gamma(1-b)}{\Gamma(1+b)}\right)
    \frac{K^c(0,0) + \tan(\pi b/2)}{K^c(0,0) - \tan(\pi b/2)} \;,
\label{eq:a0sKc}
\end{equation}
where $b=1/(n-2)$. It reduces to, for $n=4$, 
\begin{equation}
a_{l=0}/\beta_4 = \frac{K^c(0,0) + 1}{K^c(0,0) - 1} \;.
\label{eq:a0sKc4}
\end{equation}

We note that in the context of the effective 
range theory \cite{sch47,bla49,bet49,oma61},
the three parameters, $a_{gl=0}$, $a_{ul=0}$, and $\alpha_1$,
can only be expected to describe ion-atom interaction
in the ultracold regime as characterized by $\epsilon\ll s_E$,
in which only the $s$ wave makes a significant contribution.
The QDT for ion-atom interaction asserts that the
very same set of parameters can in fact describe
ion-atom interaction over a much wider range of energies,
of the order of $10^5s_E$, as tested in the next section for 
$^{23}$Na and expected to be
qualitatively the same for all alkali-metal atoms.

The two equivalent parametrizations are complementary in terms
of the physical understanding that they provide.
The parametrization using $K^c_g$, $K^c_u$, and $\alpha_1$
gives a more direct insight as to why it works over a wide
range of energies. It is because $K^c_g$ and $K^c_u$ are 
short-range parameters that are both insensitive to $\epsilon$
and $l$ \cite{gao01,gao08a,gao10a}.
The parametrization using $a_{gl=0}$, $a_{ul=0}$, and $\alpha_1$
enforces the concept that the understanding of ultracold interaction
immediately provides understanding of interactions over 
a much wider range of energies through QDT. 
This is because that embedded in the knowledge of the 
scattering lengths, $a_{gl=0}$ and $a_{ul=0}$, are the knowledge
of the $K^c_g$ and $K^c_u$ parameters, through Eq.~(\ref{eq:a0sKc}). 

\section{The example of $\mathrm{Na^++Na}$}
\label{sec:Na}

Low energy Na$^+$+Na charge exchange for $^{23}$Na has been studied
in detail by C\^ot\'e and Dalgarno in Ref.~\cite{cot00},
within the elastic approximation. 
It serves as a prototypical system to test the QDT formulation
for resonant charge exchange,
in particular the range of validity of the three-parameter
description.

In Sec.~\ref{sec:preliminary}, we make a preliminary evaluation
of the QDT description by showing, visually, 
that a three-parameter QDT parametrization,
using parameters as given in Ref.~\cite{cot00},
reproduces the cross sections of Ref.~\cite{cot00},
including all the resonance structures, 
without any knowledge of the short-range potential.
A more detailed comparison is not possible since 
Ref.~\cite{cot00} made use of unpublished potential
energy results by Magnier \textit{et al.} that are unavailable
to us.

For a more detailed comparison between fully quantum numerical 
calculations and three-parameter QDT results,
we construct in Sec.~\ref{sec:detailed} the $^2\Sigma^+_g$
and $^2\Sigma^+_u$ potential curves  for $^{23}$Na$_2^+$ 
using the same procedure as prescribed in Ref.~\cite{cot00},
except by using the later published results of
Magnier \textit{et al.} \cite{magnier96} in an intermediate region.
These potentials are meant to be as close to those
of Ref.~\cite{cot00} as possible. Fully quantum numerical
calculations of cross sections are carried out with
these potentials and compared to the results of 
corresponding three-parameter QDT descriptions,
and also to the earlier results of C\^ot\'e and Dalgarno \cite{cot00}.

In both sets of calculations, we take the sodium static
dipole polarizability to be $\alpha_1=162.7$ a.u. \cite{eks95},
the same as that adopted by C\^ot\'e and Dalgarno \cite{cot00}.
The corresponding length scale for $^{23}$Na$_2^+$ 
is $\beta_4 = (2\mu C_4/\hbar^2)^{1/2} = 1846 \;a_0$, 
where $a_0$ is the Bohr radius,
and the corresponding energy scale is
$s_E/k_B = 2.21$ $\mu$K or $s_E/h = 46.05$ kHz.

\subsection{Comparison of QDT results with previous results}
\label{sec:preliminary}

\begin{figure}
\includegraphics[width=\columnwidth]{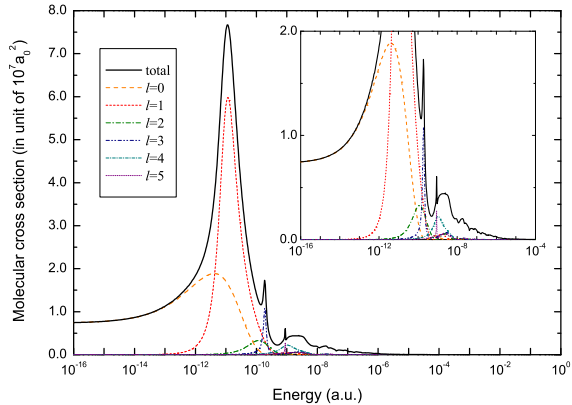}
\caption{(Color online) The total and the partial ``molecular'' cross sections for 
the $^2\Sigma^+_g$ state of $^{23}$Na$_2^+$ from 
a three-parameter QDT calculation using 
$a_{gl=0} = 763.3a_0$, $a_{ul=0} = 7721.4a_0$,
and $\alpha_1=162.7$ a.u., all from Ref.~\cite{cot00}. It is to be compared
with Fig.~2 of Ref.~\cite{cot00}. The total cross section
is obtained by summing over all partial waves until convergence.
\label{fig:coteg}}
\end{figure}

\begin{figure}
\includegraphics[width=\columnwidth]{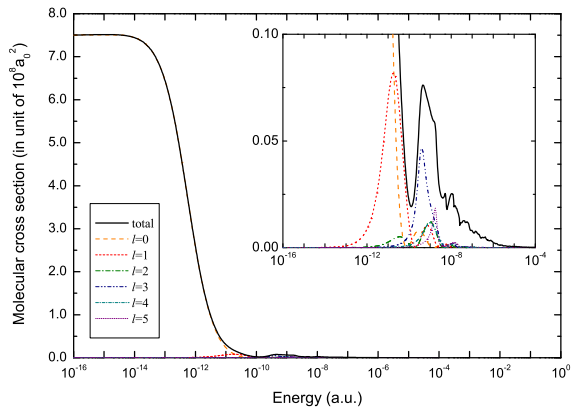}
\caption{(Color online) The same as Fig.~\ref{fig:coteg} except that it is for the 
$^2\Sigma^+_u$ state of $^{23}$Na$_2^+$. This figure is to be compared with
Fig.~3 of Ref.~\cite{cot00}. \label{fig:coteu}}
\end{figure}

Numerical results of the ``molecular'' cross sections for $^2\Sigma^+_g$
and $^2\Sigma^+_u$ states, as defined by Eq.~(\ref{eq:sigmol}), 
are given for energies ranging from $10^{-16}$ a.u.
to $1$ a.u. in Ref.~\cite{cot00}. The reference also gives the zero
energy $s$ wave scattering lengths for $^2\Sigma^+_g$ and
$^2\Sigma^+_u$ states,
\begin{subequations}
\label{eq:a0cot}
\begin{eqnarray}
a_{gl=0} &=& 763.3a_0 \;, \\
a_{ul=0} &=& 7721.4a_0 \;.
\end{eqnarray}
\end{subequations}
These $s$ wave scattering lengths,
plus the Na polarizability of $\alpha_1 = 162.7$ a.u. \cite{eks95},
give us all the parameters required for a three-parameter QDT
description of resonant charge exchange, from which all  
relevant cross sections can be calculated, without detailed 
knowledge of the potentials.

Specifically, from the $s$ wave scattering lengths of Eq.~(\ref{eq:a0cot}),
we first calculate, using Eq.~(\ref{eq:a0sKc4}), the short range $K^c$
parameters $K^c_g$ and $K^c_u$, and obtain
\begin{subequations}
\label{eq:Kccot}
\begin{eqnarray}
K^c_g &=& -2.4095 \;, \\
K^c_u &=& 1.6286 \;.
\end{eqnarray}
\end{subequations}
In the three-parameter QDT description,
they are taken as constants applicable at all energies and for all partial
waves. These parameters, together with the QDT equation for the phase
shift, Eq.~(\ref{eq:qdtpe}), give us all phase shifts and all cross
sections.
The results for the total and partial ``molecular'' cross sections
for the $^2\Sigma^+_g$ and the $^2\Sigma^+_u$ states are illustrated
in Figs.~\ref{fig:coteg} and \ref{fig:coteu}, respectively. 
They are visually nearly identical to the corresponding results shown in
Figs.~2 and 3 of Ref.~\cite{cot00}. 
All detailed features of the cross 
sections that are visible on the figures are found to be at the right
places judged by visual examination. 
The results show, at least tentatively, that the three-parameter
QDT description can provide an accurate account of resonant charge
exchange over a wide range of energies, including all the complex
structures which in this case are shape resonances from a
wide range of partial waves.

\subsection{More detailed comparison of QDT and numerical results}
\label{sec:detailed}

For a more detailed comparison between fully quantum numerical 
results and three-parameter QDT calculations,
we construct here a version of the $^2\Sigma^+_g$
and $^2\Sigma^+_u$ potential curves for Na$_2^+$.
Numerical results are calculated using these potentials and
compared to the QDT results corresponding to the same potentials.

\subsubsection{The potential energy curves adopted}

\begin{figure}
\includegraphics[width=\columnwidth]{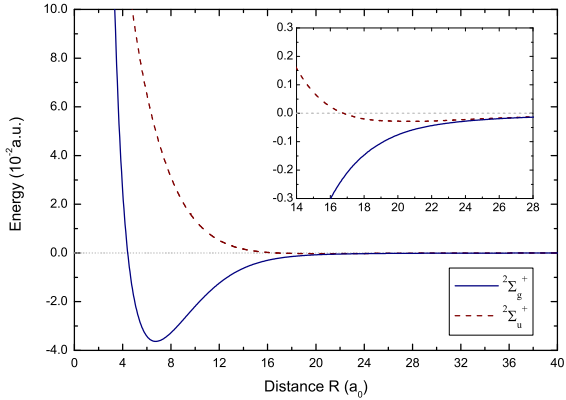}
\caption{(Color online) Potential energy curves for $^2\Sigma^+_g$ and
$^2\Sigma^+_u$ states of Na$_2^+$ adopted in our numerical
calculations. It can be compared to Fig.~1 of Ref.~\cite{cot00}. 
\label{fig:pot}}
\end{figure}

For both $^2\Sigma^+_g$ and $^2\Sigma^+_u$ states of Na$_2^+$,
we use the \textit{ab initio} data of Magnier \textit{et al.}
\cite{magnier96} ranging from $5.0\:a_0$ to $20.0\:a_0$. 
Outside of this region, the potentials are extended using the
same procedure as prescribed in Ref.~\cite{cot00},
in the hope of getting potentials as close to those of 
Ref.~\cite{cot00} as possible.
Specifically, for distance larger than $22.0\:a_0$, 
we extended the potential by the
asymptotic form  of \cite{cot00}
\begin{equation}
V_{g,u}(R) = V_{\mathrm{disp}}(R) \mp V_{\mathrm{exch}}(R) \;,
\label{eq:pot1}
\end{equation}
with $\mp$ for $^2\Sigma^+_g$ and $^2\Sigma^+_u$, respectively. The
dispersion term and exchange term are given by \cite{cot00}
\begin{eqnarray}
V_{\mathrm{disp}}(R) &=& -\frac{C_4}{R^4}
	-\frac{C_6}{R^6}-\frac{C_8}{R^8} \;, 
\label{eq:pot2} \\
V_{\mathrm{exch}}(R) &=& \frac{1}{2}AR^a e^{-b R}
\left(1+\frac{B}{R} \right) \;.
\label{eq:pot3}
\end{eqnarray}
Except for different notations \footnote{Our $C_4$, $C_6$, and $C_8$
are denoted as $C_4/2$, $C_6/2$, and $C_8/2$ in Ref.~\cite{cot00}}, 
all coefficients are taken
to be same as in Ref.~\cite{cot00}, which, in atomic units, are
given by
$C_4=\alpha_1/2=81.35$, $C_6=936.5$, $C_8=27069.5$,
$A = 0.111$, $a =2.254$, $b=0.615$, and $B = 0.494$.
Potential energies between $20.0 a_0$ and $22.0 a_0$
are interpolated using a cubic spline \cite{cppRec} 
to make a smooth connection between the \textit{ab initio} data 
and the long-range behavior. 
At short distances
($R<5.0 a_0$), we extended the potential with an
exponential wall as in Ref.~\cite{cot00}
\begin{equation}
V(R)=W\exp(-wR) \;,
\end{equation}
with
\begin{equation}
W=\left. V(R)\exp(wR) \right|_{5.0a_0}, \;\; w=\left.
-\frac{\partial \ln V(R)}{\partial R} \right|_{5.0a_0}.
\end{equation}
The resulting $^2\Sigma^+_g$ and $^2\Sigma^+_u$ potentials for Na$_2^+$,
thus constructed, are illustrated in Fig.~\ref{fig:pot}.
They have the same long-range behavior as those of Ref.~\cite{cot00},
and differ only slightly in the short range due to
slightly different \textit{ab initio} data adopted.

\subsubsection{Comparison of QDT and numerical results}

\begin{figure}
\includegraphics[width=\columnwidth]{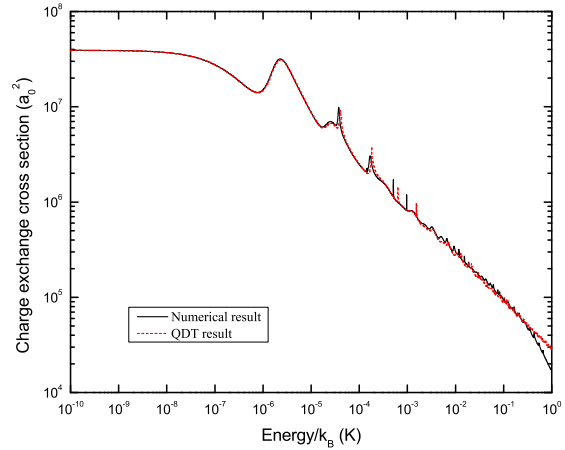}
\caption{(Color online) Comparison of the charge exchange cross sections
of $^{23}$Na obtained from a three-parameter QDT description 
(dashed line) and from numerical calculations (solid line).
\label{fig:compareEx}}
\end{figure}

\begin{figure}
\includegraphics[width=\columnwidth]{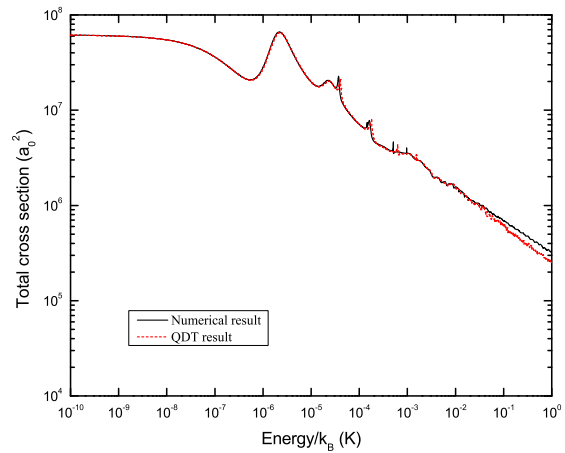}
\caption{(Color online) Comparison of the total cross sections
of $^{23}$Na$^+$+$^{23}$Na obtained from a three-parameter QDT description 
(dashed line) and from numerical calculations (solid line).
\label{fig:compareTot}}
\end{figure}

For the QDT calculations, we first calculate the
parameters $K^c_g$ and $K^c_u$, more specifically the
$K^c_g(\epsilon=0,l=0)$ and $K^c_u(\epsilon=0,l=0)$ from the potentials. 
The radial wave function is matched to
\begin{equation}
u_{\epsilon l}(r) = A_{\epsilon l}[f^c_{\epsilon_s
l}(r_s)-K^c(\epsilon,l)g^c_{\epsilon_s l}(r_s)]\;, \label{eq:kcdef}
\end{equation}
at progressively larger $R$ until the resulting $K^c$
converges to a constant to a desired accuracy \cite{gao01,gao03,gao08a}.
Here $f^c$ and $g^c$ are the zero-energy QDT reference functions for the
$-1/R^4$ potential \cite{gao04b,gao08a}.
We obtain
\begin{subequations}
\label{eq:Kcus}
\begin{eqnarray}
K^c_g &=& -1.5953 \;, \\
K^c_u &=& 0.25416 \;.
\end{eqnarray}
\end{subequations}
From the $K^c$s, the $s$ wave scattering lengths can be obtained
by substitution into Eq.~(\ref{eq:a0sKc4}). We obtain
\begin{subequations}
\label{eq:a0us}
\begin{eqnarray}
a_{gl=0} &=& 423.51a_0 \;, \\
a_{ul=0} &=& -3104.8a_0 \;. 
\end{eqnarray}
\end{subequations}
We note in passing that this method of calculating the scattering length
converges at much smaller $R$ and provides more accurate result
than by matching to the free-particle solutions or by matching
the phase shift to the effective range expansion \cite{oma61,cot00}, especially
for cases with $a_{l=0}\gg \beta_4$.

Our numerical calculations of the phase
shifts and cross sections are carried out using a 
log-derivative method \cite{joh73,mano86}.
Figure~\ref{fig:compareEx} shows the comparison of the 
charge exchange cross sections obtained from numerical
calculations and from three-parameter QDT description.
Figure~\ref{fig:compareTot} shows a similar comparison of
the total cross sections. They both show that the QDT description
and the numerical results are in excellent agreement for
energies below 0.2 mK. For energy
between 0.2 mK and 0.1 K, the QDT prediction still works well with 
the only discernible differences being due to shape resonances in high partial
waves. A better QDT description of such resonances is possible, 
but is beyond the scope of this work that focuses on the simplest
parametrization. Overall, the QDT prediction is
satisfactory below $0.1$ K, or about $10^5 s_E$. For energy
higher than $0.1$ K, the discrepancy between the two results 
grows larger. 

\begin{figure}
\includegraphics[width=\columnwidth]{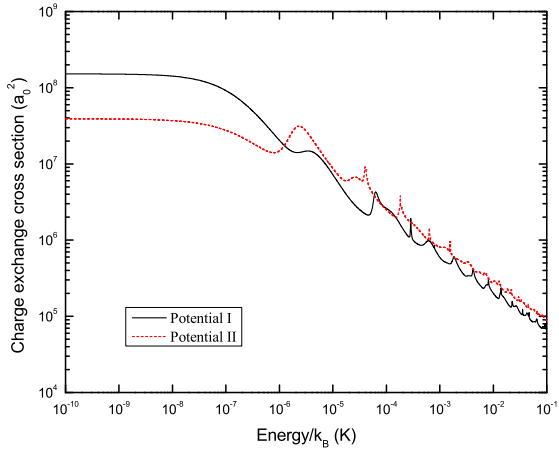}
\caption{(Color online) Comparison of charge exchange cross sections
of $^{23}$Na for two potentials both using three-parameter QDT description.
Solid line: results for the potential of Ref.~\cite{cot00}.
Dashed line: results for our potential. \label{fig:Fig6}}
\end{figure}

\begin{figure}
\includegraphics[width=\columnwidth]{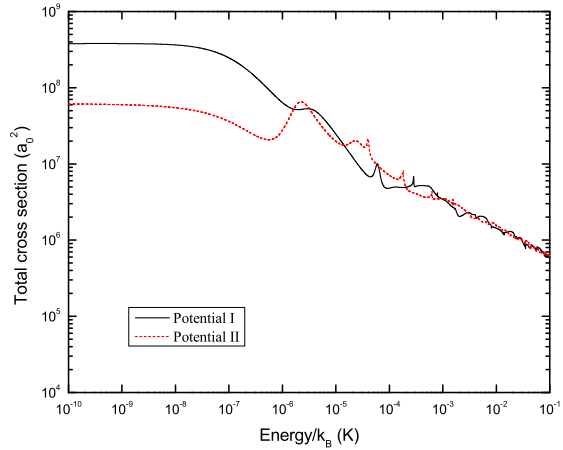}
\caption{(Color online) Comparison of total cross sections
of $^{23}$Na$^+$+$^{23}$Na obtained from three-parameter QDT descriptions
using parameters corresponding to the potential of Ref.~\cite{cot00}
(solid line) and using parameters corresponding to our 
potential (dashed line). \label{fig:Fig7}}
\end{figure}

The considerable differences between the scattering lengths for
our potentials, as given by Eq.~(\ref{eq:a0us}), and for the 
potentials of Ref.~\cite{cot00}, as given in Eq.~(\ref{eq:a0cot}), 
are illustrations of the sensitive
dependence of cold or ultracold ion-atom interactions on the
short-range potential. The two sets of potentials have the same
long-range behaviors as characterized by Eqs.~(\ref{eq:pot1})-(\ref{eq:pot3}),
and differ only slightly in the short range, to a degree that is 
visually indistinguishable on the scale of Fig.~\ref{fig:pot}.
Figures~\ref{fig:Fig6} and \ref{fig:Fig7}, which
compare QDT results for the two sets of potentials,
give a more complete picture of this dependence.
They show that the interactions in the ultracold regime 
are the most sensitive to the short-range potential.
Beyond the ultracold regime of $\epsilon\sim < s_E$, 
the shape resonance positions remain
sensitive to the potential and the QDT parameters over a considerable
wider range of energies, of the order of 1 mK or about $1000s_E$.
While the sensitive dependence of ion-atom interaction 
on the potential gradually diminishes
at higher energies for the total cross section, it remains
to a considerable degree for the charge exchange cross section.

Fortunately, the sensitive dependence of ion-atom interaction
on the PESs is fully encapsulated in a few (two) QDT parameters,
as shown in Figs.~\ref{fig:compareEx} and \ref{fig:compareTot}. 
Instead of from the PESs, this small number of parameters 
can be determined from a few experimental data points, 
such as two resonance positions for Na$^+$+Na, or two binding
energy measurements of vibrationally highly excited Na$^+_2$.
Such a determination, in a similar manner as illustrated earlier
for atom-atom interaction \cite{gao98b,gao01}, 
is further facilitated by the concept of universal
spectrum including the concept of universal resonance spectrum 
introduced in Ref.~\cite{gao10a} for the $-1/R^4$ potential.
We look forward to the availability of experimental data to 
demonstrate such applications for ion-atom systems. 

\section{Discussions and Conclusions}
\label{sec:concl}

In conclusion, we have shown that resonant charge exchange
of the type of $^1$S+$^2$S can be accurately described over a
wide range of energies using only three parameters, 
which can either be two short-range $K^c$ matrices, 
$K^c_g$ and $K^c_u$, and the atomic polarizability $\alpha_1$,
or two $s$ wave scattering lengths, $a_{gl=0}$ and $a_{ul=0}$,
and $\alpha_1$. Since the polarizability
is well known for most atoms, this is effectively a 
two-parameter description. Everything else is described by analytic
QDT functions for the $-1/R^4$ polarization potential 
(see, Ref.~\cite{gao10a} and the Appendix).
In the case of $^{23}$Na,
excellent agreement between the QDT parametrization 
and numerical results is found from 0 K 
all the way through 0.1 K, including all resonances within this range. 
To put this energy range into perspective,
we note again that $\epsilon/k_B = 0.1$ K corresponds roughly
to $\epsilon_s=\epsilon/s_E\sim 10^5$. At this energy,
one can estimate that there are at least 
$\sqrt{2}\epsilon_s^{1/4}\sim 25$ partial waves 
contributing to the cross sections. 
Such a simple parametrization over such a wide range of energies
is made possible in QDT not only by the energy insensitivity 
of the short-range parameters, $K^c_g$ and $K^c_u$,
but also by their partial-wave insensitivity \cite{gao01,gao08a,gao10a}.
The energy insensitivity is ensured here by the large length scale
separation as is typical for ion-atom interactions. 
More quantitatively, it is reflected in 
$\beta_6/\beta_4\approx 6.56\times 10^{-3}$ for Na, where
$\beta_6=(2\mu C_6/\hbar^2)^{1/4}$ is the length scale
associated with the $-C_6/R^6$ term of the potential
in Eq.~(\ref{eq:pot2}). This value gives an order-of-magnitude
measure of the length scale separation that is
representative of all alkali-metal atoms.
The partial-wave insensitivity is ensured by the combination of 
length scale separation and the smallness of the electron-to-nucleus
mass ratio \cite{gao01,gao04b}.

The example of $^{23}$Na has also served to illustrate the sensitive 
dependence of cold or ultracold ion-atom interactions on the PESs.
While one can always construct the potentials for ion-atom
systems, their accuracies
are generally far from sufficient in predicting cold
collisions, and the related vibrationally highly excited
molecular-ion spectrum \cite{gao10a}. 
The QDT deals with this difficulty by encapsulating this
sensitive dependence into a few short-range parameters
that can be determined experimentally.
The simplicity of the resulting description has 
the following implications. 
(a) Since there are only a few parameters, they can be
determined from very few experimental data points.
(b) Since the parametrization works over a wide range 
of energies, it allows the determination of the parameters
from measurements of structures (either resonance or binding
energy) away from the threshold, where they are much 
further separated \cite{gao10a} and can be
resolved with much less stringent requirement on either the
energy resolution or the temperature \cite{hec02}.
This can be an important consideration for ion-atom interactions,
where going below millikelvin has been proven to be difficult experimentally.
(c) Such a parametrization offers a systematic understanding
of a class of systems. For example, the parametrization for
Na$^+$+Na works the same for all resonant charge exchange
processes of the type of $^1$S+$^2$S. Different systems
differ only in scaling as determined by the atomic
polarizability $\alpha_1$, and the two short-range
parameters.
(d) The parameters that are used to characterize interaction
in the absence of any external field also characterize the
interaction in the presence of external fields, thus relating
field-free interactions to interactions within a field \cite{gao05a,han09}.

This work represents only the simplest QDT description for resonant
charge exchange, with a goal of establishing key qualitative features
that form the conceptual foundation for further theoretical development. 
Future works will include rigorous treatments of the hyperfine and
isotopes effects, more accurate treatment of resonances in high partial
waves which give rise to only deviations of any significance in
Figs.~~\ref{fig:compareEx} and \ref{fig:compareTot}, 
and further extension of the range of energies to room temperatures and above.
A realization of these goals, with a theory of a few parameters,
will greatly facilitate the incorporation of accurate ion-atom
interaction data into simulations of not only cold plasmas \cite{kil07}, 
but also plasma systems of interest in astrophysics and everyday 
chemistry and technology.

\begin{acknowledgments}
We thank Li You, Steven Federman, and Thomas Kvale for helpful discussions.
The work at Toledo was supported in part by NSF.
The work at Tsinghua was supported in part by the
National Science Foundation of China 
under grant number 11004116, and in part by the
research program 2010THZO of the Tsinghua University.
\end{acknowledgments}

\appendix*
\section{The $Z^{c(n)}$ matrix for $-1/R^4$ potential}

The $Z^{c(n)}$ matrix is well defined for any $-1/R^n$ type of
potentials with $n>2$ \cite{gao08a}. For the polarization potential
corresponding to $n=4$, its elements are given by  
\begin{align}
Z^{c(4)}_{fs}(\epsilon_s,l) &=
	\frac{\cos[\pi(\nu-\nu_0)/2]}
	{2M_{\epsilon_s l}\cos(\pi\nu/2)} \nonumber\\
	& \times\left\{1-(-1)^l M_{\epsilon_s l}^2\tan[\pi(\nu-\nu_0)/2]\right\} \;, 
\label{eq:Zc4fs} \\
Z^{c(4)}_{fc}(\epsilon_s,l) &=
	\frac{\cos[\pi(\nu-\nu_0)/2]}
	{2M_{\epsilon_s l}\cos(\pi\nu/2)} \nonumber\\
	& \times\left\{\tan[\pi(\nu-\nu_0)/2]-(-1)^l M_{\epsilon_s l}^2\right\} \;, \\
Z^{c(4)}_{gs}(\epsilon_s,l) &=
	\frac{\cos[\pi(\nu-\nu_0)/2]}
	{2M_{\epsilon_s l}\sin(\pi\nu/2)} \nonumber\\
	& \times\left\{1+(-1)^l M_{\epsilon_s l}^2\tan[\pi(\nu-\nu_0)/2]\right\} \;, \\
Z^{c(4)}_{gc}(\epsilon_s,l) &= 
	\frac{\cos[\pi(\nu-\nu_0)/2]}
	{2M_{\epsilon_s l}\sin(\pi\nu/2)} \nonumber\\
	& \times\left\{\tan[\pi(\nu-\nu_0)/2]+(-1)^l M_{\epsilon_s l}^2\right\} \;.
\label{eq:Zc4gc}	
\end{align}
Here $\nu_0=l+1/2$, $\nu$ is the characteristic exponent for the $-1/R^4$ potential 
(corresponding to the modified Mathieu equation \cite{abr64,hol73,khr93}), 
and $M_{\epsilon_s l}$ is one of its QDT functions. 
Their evaluations have been discussed in Ref.~\cite{gao10a}.
Together with Eqs.~(\ref{eq:Zc4fs})-(\ref{eq:Zc4gc}) 
for the $Z^{c(4)}$ matrix, they 
are all the requirements for the implementation
of QDT for resonant charge exchange.
The derivation of $Z^{c(4)}$, and other aspects of QDT for $-1/R^4$ potential, 
will be presented in a separate publication.

\bibliography{bgao,mli,eatom,twobody,coldMolion,ionChem,ionAtom,atomAtom,numerical,Rydberg,Fesh}

\begin{thebibliography}{49}%
\makeatletter
\providecommand \@ifxundefined [1]{%
 \@ifx{#1\undefined}
}%
\providecommand \@ifnum [1]{%
 \ifnum #1\expandafter \@firstoftwo
 \else \expandafter \@secondoftwo
 \fi
}%
\providecommand \@ifx [1]{%
 \ifx #1\expandafter \@firstoftwo
 \else \expandafter \@secondoftwo
 \fi
}%
\providecommand \natexlab [1]{#1}%
\providecommand \enquote  [1]{``#1''}%
\providecommand \bibnamefont  [1]{#1}%
\providecommand \bibfnamefont [1]{#1}%
\providecommand \citenamefont [1]{#1}%
\providecommand \href@noop [0]{\@secondoftwo}%
\providecommand \href [0]{\begingroup \@sanitize@url \@href}%
\providecommand \@href[1]{\@@startlink{#1}\@@href}%
\providecommand \@@href[1]{\endgroup#1\@@endlink}%
\providecommand \@sanitize@url [0]{\catcode `\\12\catcode `\$12\catcode
  `\&12\catcode `\#12\catcode `\^12\catcode `\_12\catcode `\%12\relax}%
\providecommand \@@startlink[1]{}%
\providecommand \@@endlink[0]{}%
\providecommand \url  [0]{\begingroup\@sanitize@url \@url }%
\providecommand \@url [1]{\endgroup\@href {#1}{\urlprefix }}%
\providecommand \urlprefix  [0]{URL }%
\providecommand \Eprint [0]{\href }%
\providecommand \doibase [0]{http://dx.doi.org/}%
\providecommand \selectlanguage [0]{\@gobble}%
\providecommand \bibinfo  [0]{\@secondoftwo}%
\providecommand \bibfield  [0]{\@secondoftwo}%
\providecommand \translation [1]{[#1]}%
\providecommand \BibitemOpen [0]{}%
\providecommand \bibitemStop [0]{}%
\providecommand \bibitemNoStop [0]{.\EOS\space}%
\providecommand \EOS [0]{\spacefactor3000\relax}%
\providecommand \BibitemShut  [1]{\csname bibitem#1\endcsname}%
\let\auto@bib@innerbib\@empty
\bibitem [{\citenamefont {Grier}\ \emph {et~al.}(2009)\citenamefont {Grier},
  \citenamefont {Cetina}, \citenamefont {Oru\ifmmode \check{c}\else
  \v{c}\fi{}evi\ifmmode~\acute{c}\else \'{c}\fi{}},\ and\ \citenamefont
  {Vuleti\ifmmode~\acute{c}\else \'{c}\fi{}}}]{gri09}%
  \BibitemOpen
  \bibfield  {author} {\bibinfo {author} {\bibfnamefont {A.~T.}\ \bibnamefont
  {Grier}}, \bibinfo {author} {\bibfnamefont {M.}~\bibnamefont {Cetina}},
  \bibinfo {author} {\bibfnamefont {F.}~\bibnamefont {Oru\ifmmode
  \check{c}\else \v{c}\fi{}evi\ifmmode~\acute{c}\else \'{c}\fi{}}}, \ and\
  \bibinfo {author} {\bibfnamefont {V.}~\bibnamefont
  {Vuleti\ifmmode~\acute{c}\else \'{c}\fi{}}},\ }\href {\doibase
  10.1103/PhysRevLett.102.223201} {\bibfield  {journal} {\bibinfo  {journal}
  {Phys. Rev. Lett.}\ }\textbf {\bibinfo {volume} {102}},\ \bibinfo {pages}
  {223201} (\bibinfo {year} {2009})}\BibitemShut {NoStop}%
\bibitem [{\citenamefont {Zipkes}\ \emph
  {et~al.}(2010{\natexlab{a}})\citenamefont {Zipkes}, \citenamefont {Palzer},
  \citenamefont {Sias},\ and\ \citenamefont {K\"ohl}}]{zip10}%
  \BibitemOpen
  \bibfield  {author} {\bibinfo {author} {\bibfnamefont {C.}~\bibnamefont
  {Zipkes}}, \bibinfo {author} {\bibfnamefont {S.}~\bibnamefont {Palzer}},
  \bibinfo {author} {\bibfnamefont {C.}~\bibnamefont {Sias}}, \ and\ \bibinfo
  {author} {\bibfnamefont {M.}~\bibnamefont {K\"ohl}},\ }\href@noop {}
  {\bibfield  {journal} {\bibinfo  {journal} {Nature}\ }\textbf {\bibinfo
  {volume} {464}},\ \bibinfo {pages} {388} (\bibinfo {year}
  {2010}{\natexlab{a}})}\BibitemShut {NoStop}%
\bibitem [{\citenamefont {Zipkes}\ \emph
  {et~al.}(2010{\natexlab{b}})\citenamefont {Zipkes}, \citenamefont {Palzer},
  \citenamefont {Ratschbacher}, \citenamefont {Sias},\ and\ \citenamefont
  {K\"ohl}}]{zip10b}%
  \BibitemOpen
  \bibfield  {author} {\bibinfo {author} {\bibfnamefont {C.}~\bibnamefont
  {Zipkes}}, \bibinfo {author} {\bibfnamefont {S.}~\bibnamefont {Palzer}},
  \bibinfo {author} {\bibfnamefont {L.}~\bibnamefont {Ratschbacher}}, \bibinfo
  {author} {\bibfnamefont {C.}~\bibnamefont {Sias}}, \ and\ \bibinfo {author}
  {\bibfnamefont {M.}~\bibnamefont {K\"ohl}},\ }\href {\doibase
  10.1103/PhysRevLett.105.133201} {\bibfield  {journal} {\bibinfo  {journal}
  {Phys. Rev. Lett.}\ }\textbf {\bibinfo {volume} {105}},\ \bibinfo {pages}
  {133201} (\bibinfo {year} {2010}{\natexlab{b}})}\BibitemShut {NoStop}%
\bibitem [{\citenamefont {Rellergert}\ \emph {et~al.}(2011)\citenamefont
  {Rellergert}, \citenamefont {Sullivan}, \citenamefont {Kotochigova},
  \citenamefont {Petrov}, \citenamefont {Chen}, \citenamefont {Schowalter},\
  and\ \citenamefont {Hudson}}]{rel11}%
  \BibitemOpen
  \bibfield  {author} {\bibinfo {author} {\bibfnamefont {W.~G.}\ \bibnamefont
  {Rellergert}}, \bibinfo {author} {\bibfnamefont {S.~T.}\ \bibnamefont
  {Sullivan}}, \bibinfo {author} {\bibfnamefont {S.}~\bibnamefont
  {Kotochigova}}, \bibinfo {author} {\bibfnamefont {A.}~\bibnamefont {Petrov}},
  \bibinfo {author} {\bibfnamefont {K.}~\bibnamefont {Chen}}, \bibinfo {author}
  {\bibfnamefont {S.~J.}\ \bibnamefont {Schowalter}}, \ and\ \bibinfo {author}
  {\bibfnamefont {E.~R.}\ \bibnamefont {Hudson}},\ }\href {\doibase
  10.1103/PhysRevLett.107.243201} {\bibfield  {journal} {\bibinfo  {journal}
  {Phys. Rev. Lett.}\ }\textbf {\bibinfo {volume} {107}},\ \bibinfo {pages}
  {243201} (\bibinfo {year} {2011})}\BibitemShut {NoStop}%
\bibitem [{\citenamefont {Hall}\ \emph {et~al.}(2011)\citenamefont {Hall},
  \citenamefont {Aymar}, \citenamefont {Bouloufa-Maafa}, \citenamefont
  {Dulieu},\ and\ \citenamefont {Willitsch}}]{hal11}%
  \BibitemOpen
  \bibfield  {author} {\bibinfo {author} {\bibfnamefont {F.~H.~J.}\
  \bibnamefont {Hall}}, \bibinfo {author} {\bibfnamefont {M.}~\bibnamefont
  {Aymar}}, \bibinfo {author} {\bibfnamefont {N.}~\bibnamefont
  {Bouloufa-Maafa}}, \bibinfo {author} {\bibfnamefont {O.}~\bibnamefont
  {Dulieu}}, \ and\ \bibinfo {author} {\bibfnamefont {S.}~\bibnamefont
  {Willitsch}},\ }\href {\doibase 10.1103/PhysRevLett.107.243202} {\bibfield
  {journal} {\bibinfo  {journal} {Phys. Rev. Lett.}\ }\textbf {\bibinfo
  {volume} {107}},\ \bibinfo {pages} {243202} (\bibinfo {year}
  {2011})}\BibitemShut {NoStop}%
\bibitem [{\citenamefont {Killian}(2007)}]{kil07}%
  \BibitemOpen
  \bibfield  {author} {\bibinfo {author} {\bibfnamefont {T.~C.}\ \bibnamefont
  {Killian}},\ }\href {\doibase 10.1126/science.1130556} {\bibfield  {journal}
  {\bibinfo  {journal} {Science}\ }\textbf {\bibinfo {volume} {316}},\ \bibinfo
  {pages} {705} (\bibinfo {year} {2007})}\BibitemShut {NoStop}%
\bibitem [{\citenamefont {Idziaszek}\ \emph {et~al.}(2009)\citenamefont
  {Idziaszek}, \citenamefont {Calarco}, \citenamefont {Julienne},\ and\
  \citenamefont {Simoni}}]{idz09}%
  \BibitemOpen
  \bibfield  {author} {\bibinfo {author} {\bibfnamefont {Z.}~\bibnamefont
  {Idziaszek}}, \bibinfo {author} {\bibfnamefont {T.}~\bibnamefont {Calarco}},
  \bibinfo {author} {\bibfnamefont {P.~S.}\ \bibnamefont {Julienne}}, \ and\
  \bibinfo {author} {\bibfnamefont {A.}~\bibnamefont {Simoni}},\ }\href
  {\doibase 10.1103/PhysRevA.79.010702} {\bibfield  {journal} {\bibinfo
  {journal} {Phys. Rev. A}\ }\textbf {\bibinfo {volume} {79}},\ \bibinfo {eid}
  {010702(R)} (\bibinfo {year} {2009})}\BibitemShut {NoStop}%
\bibitem [{\citenamefont {Gao}(2010{\natexlab{a}})}]{gao10a}%
  \BibitemOpen
  \bibfield  {author} {\bibinfo {author} {\bibfnamefont {B.}~\bibnamefont
  {Gao}},\ }\href {\doibase 10.1103/PhysRevLett.104.213201} {\bibfield
  {journal} {\bibinfo  {journal} {Phys. Rev. Lett.}\ }\textbf {\bibinfo
  {volume} {104}},\ \bibinfo {pages} {213201} (\bibinfo {year}
  {2010}{\natexlab{a}})}\BibitemShut {NoStop}%
\bibitem [{\citenamefont {Idziaszek}\ \emph {et~al.}(2011)\citenamefont
  {Idziaszek}, \citenamefont {Simoni}, \citenamefont {Calarco},\ and\
  \citenamefont {Julienne}}]{idz11}%
  \BibitemOpen
  \bibfield  {author} {\bibinfo {author} {\bibfnamefont {Z.}~\bibnamefont
  {Idziaszek}}, \bibinfo {author} {\bibfnamefont {A.}~\bibnamefont {Simoni}},
  \bibinfo {author} {\bibfnamefont {T.}~\bibnamefont {Calarco}}, \ and\
  \bibinfo {author} {\bibfnamefont {P.~S.}\ \bibnamefont {Julienne}},\ }\href
  {http://stacks.iop.org/1367-2630/13/i=8/a=083005} {\bibfield  {journal}
  {\bibinfo  {journal} {New Journal of Physics}\ }\textbf {\bibinfo {volume}
  {13}},\ \bibinfo {pages} {083005} (\bibinfo {year} {2011})}\BibitemShut
  {NoStop}%
\bibitem [{\citenamefont {C\^ot\'e}\ and\ \citenamefont
  {Dalgarno}(2000)}]{cot00}%
  \BibitemOpen
  \bibfield  {author} {\bibinfo {author} {\bibfnamefont {R.}~\bibnamefont
  {C\^ot\'e}}\ and\ \bibinfo {author} {\bibfnamefont {A.}~\bibnamefont
  {Dalgarno}},\ }\href {\doibase 10.1103/PhysRevA.62.012709} {\bibfield
  {journal} {\bibinfo  {journal} {Phys. Rev. A}\ }\textbf {\bibinfo {volume}
  {62}},\ \bibinfo {pages} {012709} (\bibinfo {year} {2000})}\BibitemShut
  {NoStop}%
\bibitem [{\citenamefont {Bodo}\ \emph {et~al.}(2008)\citenamefont {Bodo},
  \citenamefont {Zhang},\ and\ \citenamefont {Dalgarno}}]{Bodo2008}%
  \BibitemOpen
  \bibfield  {author} {\bibinfo {author} {\bibfnamefont {E.}~\bibnamefont
  {Bodo}}, \bibinfo {author} {\bibfnamefont {P.}~\bibnamefont {Zhang}}, \ and\
  \bibinfo {author} {\bibfnamefont {A.}~\bibnamefont {Dalgarno}},\ }\href
  {http://stacks.iop.org/1367-2630/10/i=3/a=033024} {\bibfield  {journal}
  {\bibinfo  {journal} {New Journal of Physics}\ }\textbf {\bibinfo {volume}
  {10}},\ \bibinfo {pages} {033024} (\bibinfo {year} {2008})}\BibitemShut
  {NoStop}%
\bibitem [{\citenamefont {Zhang}\ \emph {et~al.}(2009)\citenamefont {Zhang},
  \citenamefont {Dalgarno},\ and\ \citenamefont {C\^ot\'e}}]{zha09}%
  \BibitemOpen
  \bibfield  {author} {\bibinfo {author} {\bibfnamefont {P.}~\bibnamefont
  {Zhang}}, \bibinfo {author} {\bibfnamefont {A.}~\bibnamefont {Dalgarno}}, \
  and\ \bibinfo {author} {\bibfnamefont {R.}~\bibnamefont {C\^ot\'e}},\ }\href
  {\doibase 10.1103/PhysRevA.80.030703} {\bibfield  {journal} {\bibinfo
  {journal} {Phys. Rev. A}\ }\textbf {\bibinfo {volume} {80}},\ \bibinfo
  {pages} {030703(R)} (\bibinfo {year} {2009})}\BibitemShut {NoStop}%
\bibitem [{\citenamefont {Mott}\ and\ \citenamefont {Massey}(1965)}]{mot65}%
  \BibitemOpen
  \bibfield  {author} {\bibinfo {author} {\bibfnamefont {N.~F.}\ \bibnamefont
  {Mott}}\ and\ \bibinfo {author} {\bibfnamefont {H.~S.~W.}\ \bibnamefont
  {Massey}},\ }\href@noop {} {\emph {\bibinfo {title} {The Theory of Atomic
  Collisions}}}\ (\bibinfo  {publisher} {Oxford University Press, London},\
  \bibinfo {year} {1965})\BibitemShut {NoStop}%
\bibitem [{\citenamefont {Gao}(1996)}]{gao96}%
  \BibitemOpen
  \bibfield  {author} {\bibinfo {author} {\bibfnamefont {B.}~\bibnamefont
  {Gao}},\ }\href@noop {} {\bibfield  {journal} {\bibinfo  {journal} {Phys.
  Rev. A}\ }\textbf {\bibinfo {volume} {54}},\ \bibinfo {pages} {2022}
  (\bibinfo {year} {1996})}\BibitemShut {NoStop}%
\bibitem [{\citenamefont {Chin}\ \emph {et~al.}(2010)\citenamefont {Chin},
  \citenamefont {Grimm}, \citenamefont {Julienne},\ and\ \citenamefont
  {Tiesinga}}]{chi10}%
  \BibitemOpen
  \bibfield  {author} {\bibinfo {author} {\bibfnamefont {C.}~\bibnamefont
  {Chin}}, \bibinfo {author} {\bibfnamefont {R.}~\bibnamefont {Grimm}},
  \bibinfo {author} {\bibfnamefont {P.}~\bibnamefont {Julienne}}, \ and\
  \bibinfo {author} {\bibfnamefont {E.}~\bibnamefont {Tiesinga}},\ }\href
  {\doibase 10.1103/RevModPhys.82.1225} {\bibfield  {journal} {\bibinfo
  {journal} {Rev. Mod. Phys.}\ }\textbf {\bibinfo {volume} {82}},\ \bibinfo
  {pages} {1225} (\bibinfo {year} {2010})}\BibitemShut {NoStop}%
\bibitem [{\citenamefont {Gao}(2010{\natexlab{b}})}]{gao10b}%
  \BibitemOpen
  \bibfield  {author} {\bibinfo {author} {\bibfnamefont {B.}~\bibnamefont
  {Gao}},\ }\href {\doibase 10.1103/PhysRevLett.105.263203} {\bibfield
  {journal} {\bibinfo  {journal} {Phys. Rev. Lett.}\ }\textbf {\bibinfo
  {volume} {105}},\ \bibinfo {pages} {263203} (\bibinfo {year}
  {2010}{\natexlab{b}})}\BibitemShut {NoStop}%
\bibitem [{\citenamefont {Gao}(2011)}]{gao11a}%
  \BibitemOpen
  \bibfield  {author} {\bibinfo {author} {\bibfnamefont {B.}~\bibnamefont
  {Gao}},\ }\href {\doibase 10.1103/PhysRevA.83.062712} {\bibfield  {journal}
  {\bibinfo  {journal} {Phys. Rev. A}\ }\textbf {\bibinfo {volume} {83}},\
  \bibinfo {pages} {062712} (\bibinfo {year} {2011})}\BibitemShut {NoStop}%
\bibitem [{\citenamefont {Ferber}\ \emph {et~al.}(2009)\citenamefont {Ferber},
  \citenamefont {Klincare}, \citenamefont {Nikolayeva}, \citenamefont
  {Tamanis}, \citenamefont {Kn\"ockel}, \citenamefont {Tiemann},\ and\
  \citenamefont {Pashov}}]{Ferber2009}%
  \BibitemOpen
  \bibfield  {author} {\bibinfo {author} {\bibfnamefont {R.}~\bibnamefont
  {Ferber}}, \bibinfo {author} {\bibfnamefont {I.}~\bibnamefont {Klincare}},
  \bibinfo {author} {\bibfnamefont {O.}~\bibnamefont {Nikolayeva}}, \bibinfo
  {author} {\bibfnamefont {M.}~\bibnamefont {Tamanis}}, \bibinfo {author}
  {\bibfnamefont {H.}~\bibnamefont {Kn\"ockel}}, \bibinfo {author}
  {\bibfnamefont {E.}~\bibnamefont {Tiemann}}, \ and\ \bibinfo {author}
  {\bibfnamefont {A.}~\bibnamefont {Pashov}},\ }\href {\doibase
  10.1103/PhysRevA.80.062501} {\bibfield  {journal} {\bibinfo  {journal} {Phys.
  Rev. A}\ }\textbf {\bibinfo {volume} {80}},\ \bibinfo {pages} {062501}
  (\bibinfo {year} {2009})}\BibitemShut {NoStop}%
\bibitem [{\citenamefont {Knoop}\ \emph {et~al.}(2011)\citenamefont {Knoop},
  \citenamefont {Schuster}, \citenamefont {Scelle}, \citenamefont {Trautmann},
  \citenamefont {Appmeier}, \citenamefont {Oberthaler}, \citenamefont
  {Tiesinga},\ and\ \citenamefont {Tiemann}}]{kno11}%
  \BibitemOpen
  \bibfield  {author} {\bibinfo {author} {\bibfnamefont {S.}~\bibnamefont
  {Knoop}}, \bibinfo {author} {\bibfnamefont {T.}~\bibnamefont {Schuster}},
  \bibinfo {author} {\bibfnamefont {R.}~\bibnamefont {Scelle}}, \bibinfo
  {author} {\bibfnamefont {A.}~\bibnamefont {Trautmann}}, \bibinfo {author}
  {\bibfnamefont {J.}~\bibnamefont {Appmeier}}, \bibinfo {author}
  {\bibfnamefont {M.~K.}\ \bibnamefont {Oberthaler}}, \bibinfo {author}
  {\bibfnamefont {E.}~\bibnamefont {Tiesinga}}, \ and\ \bibinfo {author}
  {\bibfnamefont {E.}~\bibnamefont {Tiemann}},\ }\href {\doibase
  10.1103/PhysRevA.83.042704} {\bibfield  {journal} {\bibinfo  {journal} {Phys.
  Rev. A}\ }\textbf {\bibinfo {volume} {83}},\ \bibinfo {pages} {042704}
  (\bibinfo {year} {2011})}\BibitemShut {NoStop}%
\bibitem [{\citenamefont {Hechtfischer}\ \emph {et~al.}(2002)\citenamefont
  {Hechtfischer}, \citenamefont {Williams}, \citenamefont {Lange},
  \citenamefont {Linkemann}, \citenamefont {Schwalm}, \citenamefont {Wester},
  \citenamefont {Wolf},\ and\ \citenamefont {Zajfman}}]{hec02}%
  \BibitemOpen
  \bibfield  {author} {\bibinfo {author} {\bibfnamefont {U.}~\bibnamefont
  {Hechtfischer}}, \bibinfo {author} {\bibfnamefont {C.~J.}\ \bibnamefont
  {Williams}}, \bibinfo {author} {\bibfnamefont {M.}~\bibnamefont {Lange}},
  \bibinfo {author} {\bibfnamefont {J.}~\bibnamefont {Linkemann}}, \bibinfo
  {author} {\bibfnamefont {D.}~\bibnamefont {Schwalm}}, \bibinfo {author}
  {\bibfnamefont {R.}~\bibnamefont {Wester}}, \bibinfo {author} {\bibfnamefont
  {A.}~\bibnamefont {Wolf}}, \ and\ \bibinfo {author} {\bibfnamefont
  {D.}~\bibnamefont {Zajfman}},\ }\href {\doibase 10.1063/1.1513459} {\bibfield
   {journal} {\bibinfo  {journal} {The Journal of Chemical Physics}\ }\textbf
  {\bibinfo {volume} {117}},\ \bibinfo {pages} {8754} (\bibinfo {year}
  {2002})}\BibitemShut {NoStop}%
\bibitem [{\citenamefont {Hudson}(2009)}]{HudsonPRA2009}%
  \BibitemOpen
  \bibfield  {author} {\bibinfo {author} {\bibfnamefont {E.~R.}\ \bibnamefont
  {Hudson}},\ }\href {\doibase 10.1103/PhysRevA.79.032716} {\bibfield
  {journal} {\bibinfo  {journal} {Phys. Rev. A}\ }\textbf {\bibinfo {volume}
  {79}},\ \bibinfo {pages} {032716} (\bibinfo {year} {2009})}\BibitemShut
  {NoStop}%
\bibitem [{\citenamefont {Nguyen}\ \emph {et~al.}(2011)\citenamefont {Nguyen},
  \citenamefont {Viteri}, \citenamefont {Hohenstein}, \citenamefont {Sherrill},
  \citenamefont {Brown},\ and\ \citenamefont {Odom}}]{NguyenNJP2011}%
  \BibitemOpen
  \bibfield  {author} {\bibinfo {author} {\bibfnamefont {J.~H.~V.}\
  \bibnamefont {Nguyen}}, \bibinfo {author} {\bibfnamefont {C.~R.}\
  \bibnamefont {Viteri}}, \bibinfo {author} {\bibfnamefont {E.~G.}\
  \bibnamefont {Hohenstein}}, \bibinfo {author} {\bibfnamefont {C.~D.}\
  \bibnamefont {Sherrill}}, \bibinfo {author} {\bibfnamefont {K.~R.}\
  \bibnamefont {Brown}}, \ and\ \bibinfo {author} {\bibfnamefont {B.~C.}\
  \bibnamefont {Odom}},\ }\href
  {http://stacks.iop.org/1367-2630/13/i=6/a=063023} {\bibfield  {journal}
  {\bibinfo  {journal} {New Journal of Physics}\ }\textbf {\bibinfo {volume}
  {13}},\ \bibinfo {pages} {063023} (\bibinfo {year} {2011})}\BibitemShut
  {NoStop}%
\bibitem [{\citenamefont {Nguyen}\ and\ \citenamefont
  {Odom}(2011)}]{NguyenPRA2011}%
  \BibitemOpen
  \bibfield  {author} {\bibinfo {author} {\bibfnamefont {J.~H.~V.}\
  \bibnamefont {Nguyen}}\ and\ \bibinfo {author} {\bibfnamefont
  {B.}~\bibnamefont {Odom}},\ }\href {\doibase 10.1103/PhysRevA.83.053404}
  {\bibfield  {journal} {\bibinfo  {journal} {Phys. Rev. A}\ }\textbf {\bibinfo
  {volume} {83}},\ \bibinfo {pages} {053404} (\bibinfo {year}
  {2011})}\BibitemShut {NoStop}%
\bibitem [{\citenamefont {Willitsch}(0)}]{wil12}%
  \BibitemOpen
  \bibfield  {author} {\bibinfo {author} {\bibfnamefont {S.}~\bibnamefont
  {Willitsch}},\ }\href {\doibase 10.1080/0144235X.2012.667221} {\bibfield
  {journal} {\bibinfo  {journal} {International Reviews in Physical Chemistry}\
  }\textbf {\bibinfo {volume} {0}},\ \bibinfo {pages} {1} (\bibinfo {year}
  {0})}\BibitemShut {NoStop}%
\bibitem [{\citenamefont {Igarashi}\ and\ \citenamefont {Lin}(1999)}]{iga99}%
  \BibitemOpen
  \bibfield  {author} {\bibinfo {author} {\bibfnamefont {A.}~\bibnamefont
  {Igarashi}}\ and\ \bibinfo {author} {\bibfnamefont {C.~D.}\ \bibnamefont
  {Lin}},\ }\href {\doibase 10.1103/PhysRevLett.83.4041} {\bibfield  {journal}
  {\bibinfo  {journal} {Phys. Rev. Lett.}\ }\textbf {\bibinfo {volume} {83}},\
  \bibinfo {pages} {4041} (\bibinfo {year} {1999})}\BibitemShut {NoStop}%
\bibitem [{\citenamefont {Esry}\ \emph {et~al.}(2000)\citenamefont {Esry},
  \citenamefont {Sadeghpour}, \citenamefont {Wells},\ and\ \citenamefont
  {Ben-Itzhak}}]{esr00}%
  \BibitemOpen
  \bibfield  {author} {\bibinfo {author} {\bibfnamefont {B.~D.}\ \bibnamefont
  {Esry}}, \bibinfo {author} {\bibfnamefont {H.~R.}\ \bibnamefont
  {Sadeghpour}}, \bibinfo {author} {\bibfnamefont {E.}~\bibnamefont {Wells}}, \
  and\ \bibinfo {author} {\bibfnamefont {I.}~\bibnamefont {Ben-Itzhak}},\
  }\href {http://stacks.iop.org/0953-4075/33/i=23/a=306} {\bibfield  {journal}
  {\bibinfo  {journal} {Journal of Physics B: Atomic, Molecular and Optical
  Physics}\ }\textbf {\bibinfo {volume} {33}},\ \bibinfo {pages} {5329}
  (\bibinfo {year} {2000})}\BibitemShut {NoStop}%
\bibitem [{\citenamefont {Watanabe}\ and\ \citenamefont
  {Greene}(1980)}]{wat80}%
  \BibitemOpen
  \bibfield  {author} {\bibinfo {author} {\bibfnamefont {S.}~\bibnamefont
  {Watanabe}}\ and\ \bibinfo {author} {\bibfnamefont {C.~H.}\ \bibnamefont
  {Greene}},\ }\href@noop {} {\bibfield  {journal} {\bibinfo  {journal} {Phys.
  Rev. A}\ }\textbf {\bibinfo {volume} {22}},\ \bibinfo {pages} {158} (\bibinfo
  {year} {1980})}\BibitemShut {NoStop}%
\bibitem [{\citenamefont {Fabrikant}(1986)}]{fab86}%
  \BibitemOpen
  \bibfield  {author} {\bibinfo {author} {\bibfnamefont {I.~I.}\ \bibnamefont
  {Fabrikant}},\ }\href@noop {} {\bibfield  {journal} {\bibinfo  {journal} {J.
  Phys. B}\ }\textbf {\bibinfo {volume} {19}},\ \bibinfo {pages} {1527}
  (\bibinfo {year} {1986})}\BibitemShut {NoStop}%
\bibitem [{\citenamefont {Gao}(2001)}]{gao01}%
  \BibitemOpen
  \bibfield  {author} {\bibinfo {author} {\bibfnamefont {B.}~\bibnamefont
  {Gao}},\ }\href@noop {} {\bibfield  {journal} {\bibinfo  {journal} {Phys.
  Rev. A}\ }\textbf {\bibinfo {volume} {64}},\ \bibinfo {pages} {010701(R)}
  (\bibinfo {year} {2001})}\BibitemShut {NoStop}%
\bibitem [{\citenamefont {Gao}(2008)}]{gao08a}%
  \BibitemOpen
  \bibfield  {author} {\bibinfo {author} {\bibfnamefont {B.}~\bibnamefont
  {Gao}},\ }\href@noop {} {\bibfield  {journal} {\bibinfo  {journal} {Phys.
  Rev. A}\ }\textbf {\bibinfo {volume} {78}},\ \bibinfo {pages} {012702}
  (\bibinfo {year} {2008})}\BibitemShut {NoStop}%
\bibitem [{\citenamefont {Gao}\ \emph {et~al.}(2005)\citenamefont {Gao},
  \citenamefont {Tiesinga}, \citenamefont {Williams},\ and\ \citenamefont
  {Julienne}}]{gao05a}%
  \BibitemOpen
  \bibfield  {author} {\bibinfo {author} {\bibfnamefont {B.}~\bibnamefont
  {Gao}}, \bibinfo {author} {\bibfnamefont {E.}~\bibnamefont {Tiesinga}},
  \bibinfo {author} {\bibfnamefont {C.~J.}\ \bibnamefont {Williams}}, \ and\
  \bibinfo {author} {\bibfnamefont {P.~S.}\ \bibnamefont {Julienne}},\
  }\href@noop {} {\bibfield  {journal} {\bibinfo  {journal} {Phys. Rev. A}\
  }\textbf {\bibinfo {volume} {72}},\ \bibinfo {pages} {042719} (\bibinfo
  {year} {2005})}\BibitemShut {NoStop}%
\bibitem [{\citenamefont {Landau}\ and\ \citenamefont
  {Lifshitz}(1977)}]{lan77}%
  \BibitemOpen
  \bibfield  {author} {\bibinfo {author} {\bibfnamefont {L.~D.}\ \bibnamefont
  {Landau}}\ and\ \bibinfo {author} {\bibfnamefont {E.~M.}\ \bibnamefont
  {Lifshitz}},\ }\href@noop {} {\emph {\bibinfo {title} {Quantum Mechanics}}}\
  (\bibinfo  {publisher} {Pergamon Press, Oxford},\ \bibinfo {year}
  {1977})\BibitemShut {NoStop}%
\bibitem [{\citenamefont {Gao}(2003)}]{gao03}%
  \BibitemOpen
  \bibfield  {author} {\bibinfo {author} {\bibfnamefont {B.}~\bibnamefont
  {Gao}},\ }\href@noop {} {\bibfield  {journal} {\bibinfo  {journal} {J. Phys.
  B}\ }\textbf {\bibinfo {volume} {36}},\ \bibinfo {pages} {2111} (\bibinfo
  {year} {2003})}\BibitemShut {NoStop}%
\bibitem [{\citenamefont {Gao}(2004)}]{gao04b}%
  \BibitemOpen
  \bibfield  {author} {\bibinfo {author} {\bibfnamefont {B.}~\bibnamefont
  {Gao}},\ }\href@noop {} {\bibfield  {journal} {\bibinfo  {journal} {Euro.
  Phys. J. D}\ }\textbf {\bibinfo {volume} {31}},\ \bibinfo {pages} {283}
  (\bibinfo {year} {2004})}\BibitemShut {NoStop}%
\bibitem [{\citenamefont {Schwinger}(1947)}]{sch47}%
  \BibitemOpen
  \bibfield  {author} {\bibinfo {author} {\bibfnamefont {J.}~\bibnamefont
  {Schwinger}},\ }\href {\doibase 10.1103/PhysRev.72.738} {\bibfield  {journal}
  {\bibinfo  {journal} {Phys. Rev.}\ }\textbf {\bibinfo {volume} {72}},\
  \bibinfo {pages} {738} (\bibinfo {year} {1947})}\BibitemShut {NoStop}%
\bibitem [{\citenamefont {Blatt}\ and\ \citenamefont {Jackson}(1949)}]{bla49}%
  \BibitemOpen
  \bibfield  {author} {\bibinfo {author} {\bibfnamefont {J.~M.}\ \bibnamefont
  {Blatt}}\ and\ \bibinfo {author} {\bibfnamefont {D.~J.}\ \bibnamefont
  {Jackson}},\ }\href@noop {} {\bibfield  {journal} {\bibinfo  {journal} {Phys.
  Rev.}\ }\textbf {\bibinfo {volume} {76}},\ \bibinfo {pages} {18} (\bibinfo
  {year} {1949})}\BibitemShut {NoStop}%
\bibitem [{\citenamefont {Bethe}(1949)}]{bet49}%
  \BibitemOpen
  \bibfield  {author} {\bibinfo {author} {\bibfnamefont {H.~A.}\ \bibnamefont
  {Bethe}},\ }\href {\doibase 10.1103/PhysRev.76.38} {\bibfield  {journal}
  {\bibinfo  {journal} {Phys. Rev.}\ }\textbf {\bibinfo {volume} {76}},\
  \bibinfo {pages} {38} (\bibinfo {year} {1949})}\BibitemShut {NoStop}%
\bibitem [{\citenamefont {O'Malley}\ \emph {et~al.}(1961)\citenamefont
  {O'Malley}, \citenamefont {Spruch},\ and\ \citenamefont {Rosenberg}}]{oma61}%
  \BibitemOpen
  \bibfield  {author} {\bibinfo {author} {\bibfnamefont {T.~F.}\ \bibnamefont
  {O'Malley}}, \bibinfo {author} {\bibfnamefont {L.}~\bibnamefont {Spruch}}, \
  and\ \bibinfo {author} {\bibfnamefont {L.}~\bibnamefont {Rosenberg}},\ }\href
  {\doibase 10.1063/1.1703735} {\bibfield  {journal} {\bibinfo  {journal} {J.
  Math. Phys.}\ }\textbf {\bibinfo {volume} {2}},\ \bibinfo {pages} {491}
  (\bibinfo {year} {1961})}\BibitemShut {NoStop}%
\bibitem [{\citenamefont {Magnier}\ and\ \citenamefont
  {Masnou-Seeuws}(1996)}]{magnier96}%
  \BibitemOpen
  \bibfield  {author} {\bibinfo {author} {\bibfnamefont {S.}~\bibnamefont
  {Magnier}}\ and\ \bibinfo {author} {\bibfnamefont {F.}~\bibnamefont
  {Masnou-Seeuws}},\ }\href@noop {} {\bibfield  {journal} {\bibinfo  {journal}
  {Mol. Phys.}\ }\textbf {\bibinfo {volume} {89}},\ \bibinfo {pages} {711}
  (\bibinfo {year} {1996})}\BibitemShut {NoStop}%
\bibitem [{\citenamefont {Ekstrom}\ \emph {et~al.}(1995)\citenamefont
  {Ekstrom}, \citenamefont {Schmiedmayer}, \citenamefont {Chapman},
  \citenamefont {Hammond},\ and\ \citenamefont {Pritchard}}]{eks95}%
  \BibitemOpen
  \bibfield  {author} {\bibinfo {author} {\bibfnamefont {C.~R.}\ \bibnamefont
  {Ekstrom}}, \bibinfo {author} {\bibfnamefont {J.}~\bibnamefont
  {Schmiedmayer}}, \bibinfo {author} {\bibfnamefont {M.~S.}\ \bibnamefont
  {Chapman}}, \bibinfo {author} {\bibfnamefont {T.~D.}\ \bibnamefont
  {Hammond}}, \ and\ \bibinfo {author} {\bibfnamefont {D.~E.}\ \bibnamefont
  {Pritchard}},\ }\href {\doibase 10.1103/PhysRevA.51.3883} {\bibfield
  {journal} {\bibinfo  {journal} {Phys. Rev. A}\ }\textbf {\bibinfo {volume}
  {51}},\ \bibinfo {pages} {3883} (\bibinfo {year} {1995})}\BibitemShut
  {NoStop}%
\bibitem [{Note1()}]{Note1}%
  \BibitemOpen
  \bibinfo {note} {Our $C_4$, $C_6$, and $C_8$ are denoted as $C_4/2$, $C_6/2$,
  and $C_8/2$ in Ref.~\cite {cot00}}\BibitemShut {NoStop}%
\bibitem [{\citenamefont {William H.~Press}\ and\ \citenamefont
  {Flannery}(2007)}]{cppRec}%
  \BibitemOpen
  \bibfield  {author} {\bibinfo {author} {\bibfnamefont {W.~T.~V.}\
  \bibnamefont {William H.~Press}, \bibfnamefont {Saul A.~Teukolsky}}\ and\
  \bibinfo {author} {\bibfnamefont {B.~P.}\ \bibnamefont {Flannery}},\
  }\href@noop {} {\emph {\bibinfo {title} {Numerical Recipes: The Art of
  Scientific Computing}}},\ \bibinfo {edition} {3rd}\ ed.\ (\bibinfo
  {publisher} {Cambridge University Press, Cambridge},\ \bibinfo {year}
  {2007})\BibitemShut {NoStop}%
\bibitem [{\citenamefont {Johnson}(1973)}]{joh73}%
  \BibitemOpen
  \bibfield  {author} {\bibinfo {author} {\bibfnamefont {B.~R.}\ \bibnamefont
  {Johnson}},\ }\href {\doibase DOI: 10.1016/0021-9991(73)90049-1} {\bibfield
  {journal} {\bibinfo  {journal} {Journal of Computational Physics}\ }\textbf
  {\bibinfo {volume} {13}},\ \bibinfo {pages} {445 } (\bibinfo {year}
  {1973})}\BibitemShut {NoStop}%
\bibitem [{\citenamefont {Manonopoulos}(1986)}]{mano86}%
  \BibitemOpen
  \bibfield  {author} {\bibinfo {author} {\bibfnamefont {D.~E.}\ \bibnamefont
  {Manonopoulos}},\ }\href@noop {} {\bibfield  {journal} {\bibinfo  {journal}
  {J. Chem. Phys.}\ }\textbf {\bibinfo {volume} {85}},\ \bibinfo {pages} {6425}
  (\bibinfo {year} {1986})}\BibitemShut {NoStop}%
\bibitem [{\citenamefont {Gao}(1998)}]{gao98b}%
  \BibitemOpen
  \bibfield  {author} {\bibinfo {author} {\bibfnamefont {B.}~\bibnamefont
  {Gao}},\ }\href@noop {} {\bibfield  {journal} {\bibinfo  {journal} {Phys.
  Rev. A}\ }\textbf {\bibinfo {volume} {58}},\ \bibinfo {pages} {4222}
  (\bibinfo {year} {1998})}\BibitemShut {NoStop}%
\bibitem [{\citenamefont {Hanna}\ \emph {et~al.}(2009)\citenamefont {Hanna},
  \citenamefont {Tiesinga},\ and\ \citenamefont {Julienne}}]{han09}%
  \BibitemOpen
  \bibfield  {author} {\bibinfo {author} {\bibfnamefont {T.~M.}\ \bibnamefont
  {Hanna}}, \bibinfo {author} {\bibfnamefont {E.}~\bibnamefont {Tiesinga}}, \
  and\ \bibinfo {author} {\bibfnamefont {P.~S.}\ \bibnamefont {Julienne}},\
  }\href@noop {} {\bibfield  {journal} {\bibinfo  {journal} {Phys. Rev. A}\
  }\textbf {\bibinfo {volume} {79}},\ \bibinfo {eid} {040701} (\bibinfo {year}
  {2009})}\BibitemShut {NoStop}%
\bibitem [{\citenamefont {Abramowitz}\ and\ \citenamefont
  {Stegun}(1964)}]{abr64}%
  \BibitemOpen
  \bibinfo {editor} {\bibfnamefont {M.}~\bibnamefont {Abramowitz}}\ and\
  \bibinfo {editor} {\bibfnamefont {I.~A.}\ \bibnamefont {Stegun}},\ eds.,\
  \href@noop {} {\emph {\bibinfo {title} {Handbook of Mathematical
  Functions}}}\ (\bibinfo  {publisher} {National Bureau of Standards,
  Washington, D.C.},\ \bibinfo {year} {1964})\BibitemShut {NoStop}%
\bibitem [{\citenamefont {Holzwarth}(1973)}]{hol73}%
  \BibitemOpen
  \bibfield  {author} {\bibinfo {author} {\bibfnamefont {N.~A.~W.}\
  \bibnamefont {Holzwarth}},\ }\href {\doibase 10.1063/1.1666295} {\bibfield
  {journal} {\bibinfo  {journal} {J. Math. Phys.}\ }\textbf {\bibinfo {volume}
  {14}},\ \bibinfo {pages} {191} (\bibinfo {year} {1973})}\BibitemShut
  {NoStop}%
\bibitem [{\citenamefont {Khrebtukov}(1993)}]{khr93}%
  \BibitemOpen
  \bibfield  {author} {\bibinfo {author} {\bibfnamefont {D.~B.}\ \bibnamefont
  {Khrebtukov}},\ }\href@noop {} {\bibfield  {journal} {\bibinfo  {journal} {J.
  Phys. A}\ }\textbf {\bibinfo {volume} {26}},\ \bibinfo {pages} {6357}
  (\bibinfo {year} {1993})}\BibitemShut {NoStop}%
\end{thebibliography}%

\end{document}